\documentclass[aps,prl
		,twocolumn
		,groupedaddress]{revtex4}

\usepackage{graphicx,hyperref}

\newcommand{\etal}{\textit{et al.}}
\newcommand{\mos}{MoS$_2$}
\newcommand{\mose}{MoSe$_2$}
\renewcommand{\i}{\textit{(i)}}
\newcommand{\ii}{\textit{(ii)}}
\newcommand{\iii}{\textit{(iii)}}
\usepackage{color}

      \definecolor{darkgreen}{RGB}{0,128,0}
	
\usepackage{overpic
	    ,paralist}

\begin{document}

\mbox{
\href{https://doi.org/10.1103/PhysRevLett.121.167401}{This article has been published at Physical Review Letters: Phys. Rev. Lett. 121, 167401 (2018).}
}

\title{Resonance Profiles of Valley Polarization in Single-Layer \mos{} and \mose{}}

\author{Hans Tornatzky}
\email{ht07@physik.tu-berlin.de}
\affiliation{Institut f\"ur Festk\"orperphysik, Technische Universit\"at Berlin, Hardenbergstraße 36, 10623 Berlin, Germany}

\author{Anne-Marie Kaulitz}
\affiliation{Institut f\"ur Festk\"orperphysik, Technische Universit\"at Berlin, Hardenbergstraße 36, 10623 Berlin, Germany}

\author{Janina Maultzsch}
\affiliation{Institut f\"ur Festk\"orperphysik, Technische Universit\"at Berlin, Hardenbergstraße 36, 10623 Berlin, Germany}
\affiliation{Department Physik, Friedrich-Alexander-Universit\"at Erlangen-N\"urnberg, Staudtstraße 7, 91058 Erlangen, Germany}

\date{\today}

  \begin{abstract}
   In this letter we present photoluminescence measurements with different excitation energies on single-layer \mos{} and \mose{} in order to examine the 
resonance behavior of the conservation of circular polarization in these transition metal dichalcogenides. 
   We find that the circular polarization of the emitted light is conserved to 100\,\% in \mos{} and 84\,\%/79\,\% (\textit{A}/$A^-$ peaks) in \mose{} close to 
resonance. The values for \mose{} surpass any previously reported value. However, in contrast to previous predictions, the degree of circular polarization 
decreases clearly at energies less than the two-phonon longitudinal acoustic phonon energy above the resonance.
   
   Our findings indicate that at least two competing processes underly the depolarization of the emission in single-layer transition metal dichalcogenides.
  \end{abstract}

\maketitle

\begin{figure}[b]
	\includegraphics{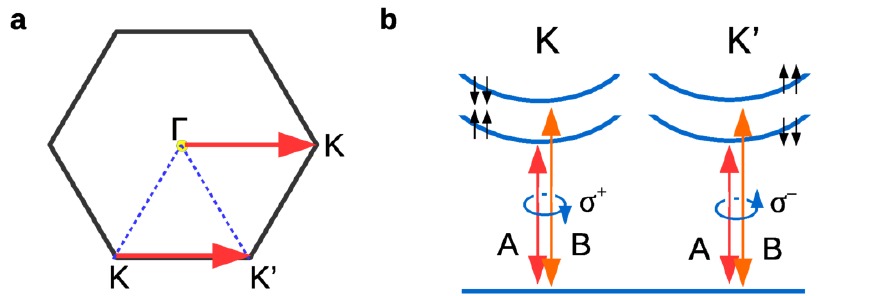}
     \caption{\textsf{
     (a) Schematic illustration of the reciprocal space of hexagonal systems; the two arrows are of the same length and orientation, indicating that \textit{K} to \textit{K'} scattering requires the momentum of a \textit{K}-point phonon. 
     (b) Illustration of the two bands contributing to the \textit{A} and \textit{B} excitonic transitions at the \textit{K} and \textit{K'} points; black arrows indicate the electron and hole spin of corresponding bands. 
     }}
     \label{fig:intro}
    \end{figure}

    Transition metal dichalcogenides (TMDCs) like \mos{} and \mose{} have emerged as promising materials for various future applications due to their 
properties when thinned down to a single layer. In the single-layer limit, they become direct semiconductors with their band gaps located at the \textit{K} points in 
the Brillouin zone [cf. Fig.~\ref{fig:intro}(a)].
    Strong spin-orbit coupling results in a large valence band splitting of a few hundred meV at the \textit{K} points; the conduction bands are split by a few 
meV. This splitting causes the so-called \textit{A} and \textit{B} exciton transitions [Fig.~\ref{fig:intro}(b)].
    Optical selection rules for circularly polarized light in single layers lead to transitions where electrons and holes are exclusively generated at either 
\textit{K} or \textit{K'}, as time reversal symmetry and the absence of inversion symmetry couple spin and valley (so-called valley 
polarization)\,\cite{Xiao2012,Xu2014, Glazov2015}. 
    Hence, in theory excitation with circularly polarized light is followed by the emission of a photon with the same circular polarization.
    However, since the first published experimental evidence of valley polarization in 2012\,\cite{Mak2012a,Zeng2012}, most groups find nonperfect 
polarization in different experimental conditions\,(see, e.g.,~\cite{Cao2012,Zeng2012,Sallen2012}). 
    The origin of such reduced circular polarization of the emitted light from these materials is, however, still controversially discussed. Different 
approaches of how the conservation of circular polarization is limited are suggested in the literature. 
    Two of these mechanisms are based on the formation of an exciton in the respective other \textit{K} point. The first, often called "2LA mechanism," is a 
two-longitudinal acoustic-phonon assisted scattering of the exciton\,\cite{KioseoglouAPL2012,KioseoglouSciRep2016,KioseoglouPSSRRL2016}. The second 
suggested process generates an exciton at \textit{K'} (\textit{K}) while annihilating the photoexcited exciton at \textit{K} (\textit{K'}) by interference of 
the exciton wave functions\,\cite{Yu2014,Glazov2014,Baranowski2017}. We will refer to this process as the "valley exchange mechanism".

    One of the reasons for the diverse speculations on the depolarization mechanism is certainly the lack of data with excitation close to the emission lines, 
in particular from within the energy range of two LA phonons above the emission energy.
   
    In this Letter we present circularly polarized photoluminescence excitation (PLE) measurements of \mos{} and \mose{} close to the resonance of their 
exciton and trion emissions. We were able to record spectra less than 5\,meV away from the excitation energy by using a triple monochromator setup, thereby 
overcoming the technical restrictions of previously published results.
    Further, we used tunable lasers to analyze the resonance, rather than detuning the emission by changing the sample temperature, as has been done in 
most previous publications. We are thereby able to decouple our observations from temperature induced effects, which are proven to be strong\,\cite{Zhu2014}.
    We find that the emitted circular polarized light is conserved to 100\,\% in \mos{} and 84\,\%/79\,\% (\textit{A}/$A^-$ peaks) in \mose{} when exciting 
close to the resonance. However, in our experiments the decrease of the degree of polarization starts at lower excess energy than the previously predicted 2LA 
phonon energy equivalent.
    Further, we observe with 84\,\% / 79\,\% the highest so far reported degree of polarization for \mose{}.
      
\begin{figure}
      \includegraphics[width=\columnwidth]{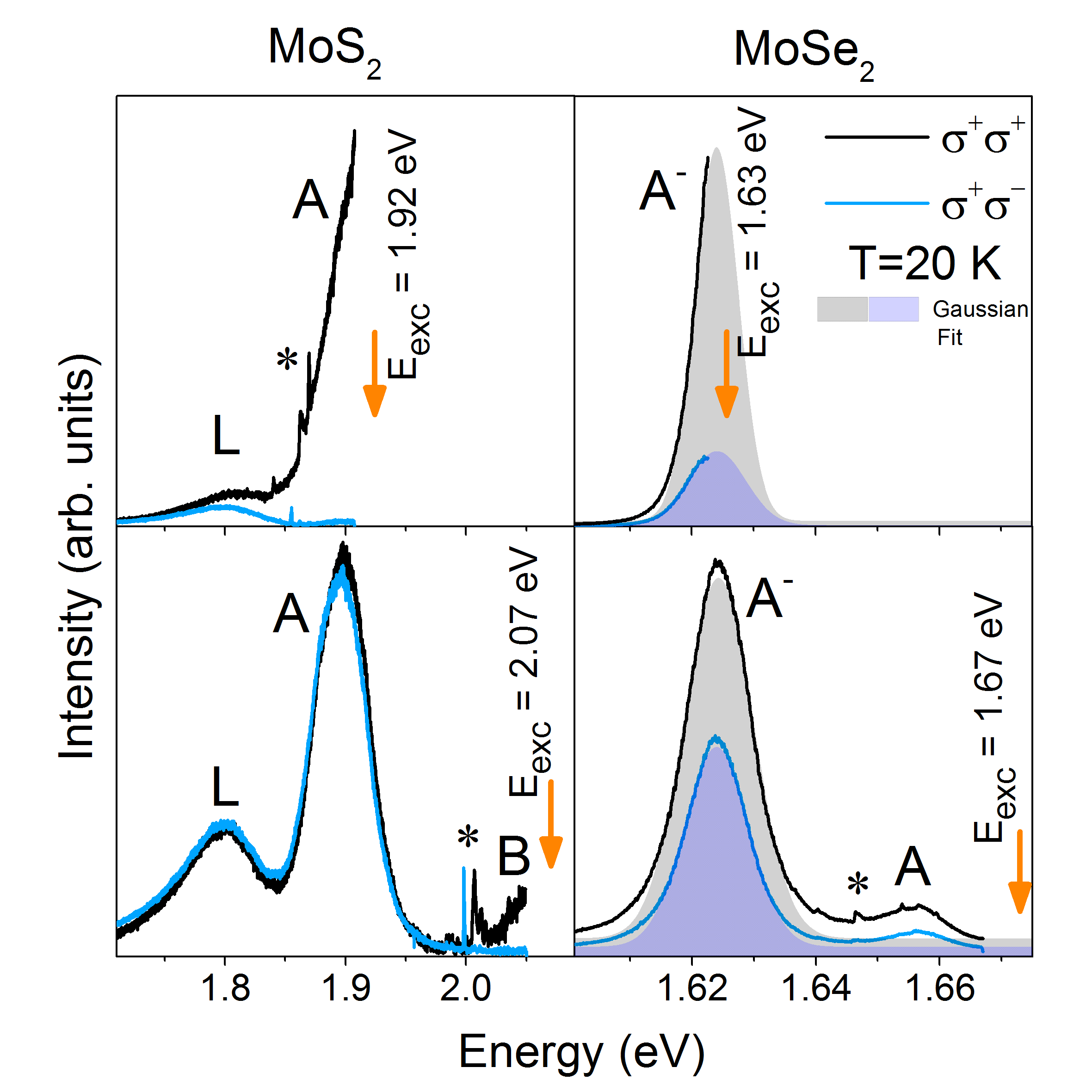}
	\put(-215,215){\bf \textsf{a}}\put(-215,118){\bf \textsf{b}}
	\put(-110,215){\bf \textsf{c}}\put(-110,118){\bf \textsf{d}}
     \caption{\textsf{
     Typical circular-polarization resolved PL spectra in resonance with the \textit{A} (a) and \textit{B} exciton (b) of \mos{} and spectra in resonance (c) 
and out of resonance (d) with the trion of \mose. Arrows indicate the excitation energies. Narrow peaks in the vicinity of asterisks (*) are due to Raman 
scattering and were excluded for fitting.
     }}
     \label{fig:PL-spectra}
    \end{figure}

      Typical circular-polarization resolved photoluminescence (PL) spectra 
      at 20\,K are shown in Fig.~\ref{fig:PL-spectra}. For \mos{}, the \textit{A} emission is found at 1.897\,eV (the trion peak emerges after 
irradiation with higher laser powers, cf.~Fig.~S2 in the Supplemental Material~\cite{SI}).
      The \textit{B} exciton emission is found at 2.060\,eV, 163\,meV above the \textit{A} peak, which is in good agreement with the combined conduction band 
and valence band spin-orbit splitting values of a few meV\,\cite{Kosmider2013} (CB) + 160\,meV\,\cite{Mak2012c} (VB). An additional peak at 1.803\,eV can 
be observed, potentially caused by localized excitons ($L$)\,\cite{Plechinger2012,Tongay2013}. Furthermore, Raman lines are superimposed on the PL emission 
(cf. Fig.~\ref{fig:PL-spectra}, marked with asterisks). 
    
      In the case of \mose{}, the emission lines of the \textit{A} and $A^-$ have a full width at half maximum smaller than the trion binding energy, 
and are therefore well separated at energies of $1.656$ and $1.624$\,eV, respectively. 
      
      To quantitatively address the degree of valley polarization, the contrastlike degree of circular polarization (DOP) 
$\rho={(I_{\sigma^+}-I_{\sigma^-})}/{(I_{\sigma^+}+I_{\sigma^-})}$ is introduced, with $I_{\sigma^+}$ ($I_{\sigma^-}$) being the photoluminescence (PL) emission 
intensity of the $\sigma^+$ ($\sigma^-$) circular polarization. 
%
%
    
      
%

%
        \begin{figure*}[htbp!]
    \includegraphics[trim= 50 20 83 50, clip, 
		      width=.98\columnwidth]{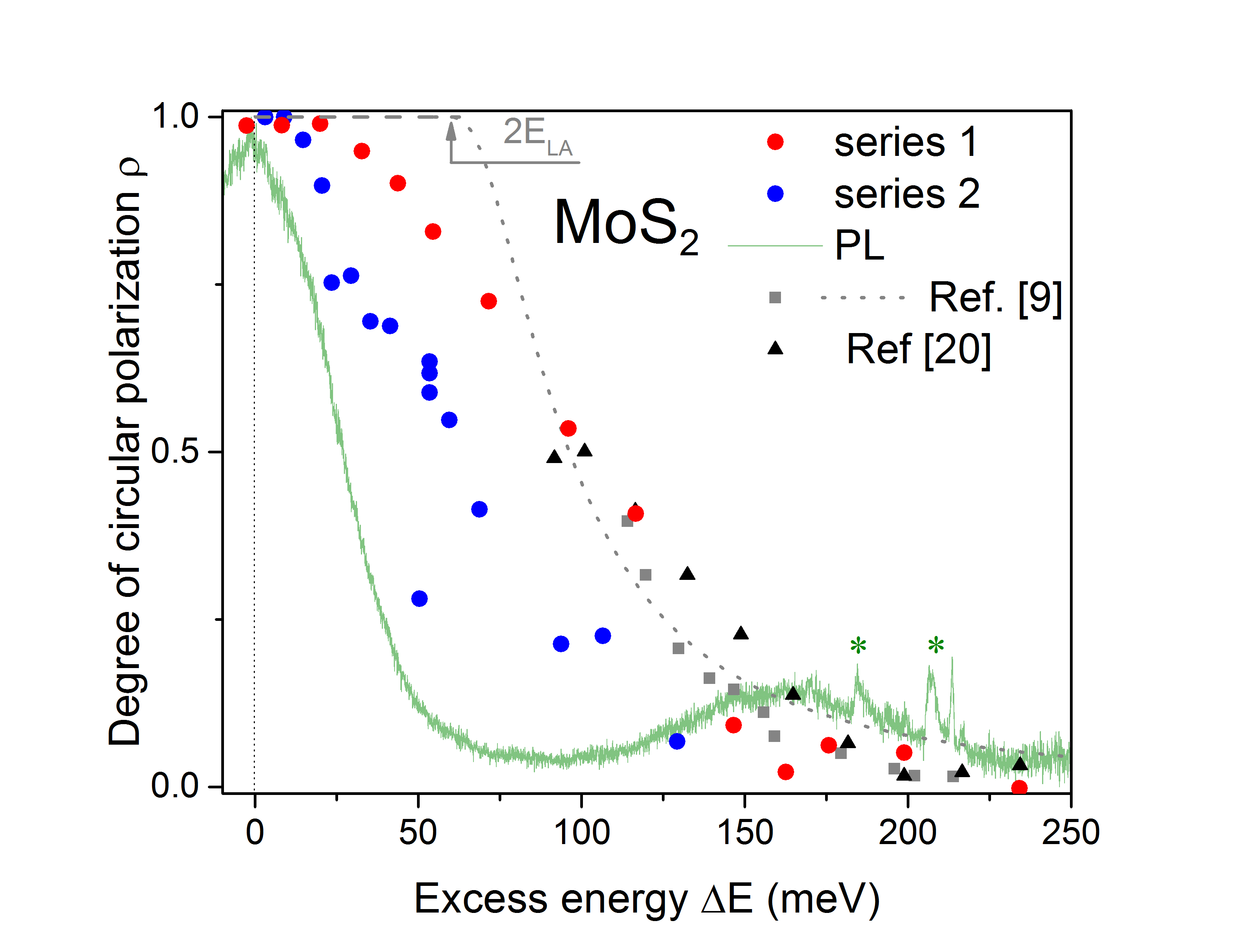}
    \put(-235,198){\bf \textsf{a}} 
    \includegraphics[trim= 50 20 83 50, clip, 
		      width=.98\columnwidth]{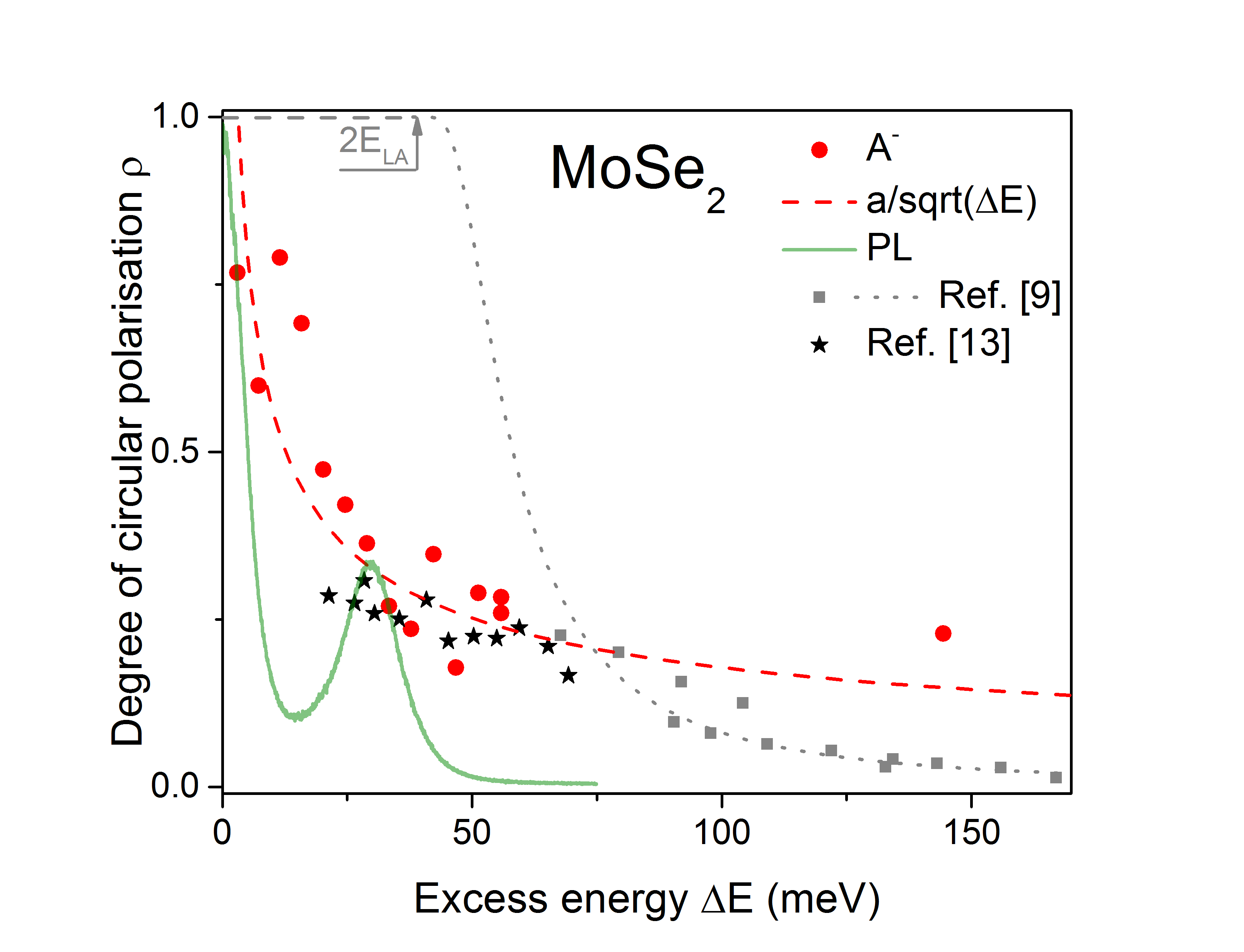}\put(-80,70){\includegraphics[width=.28\columnwidth]{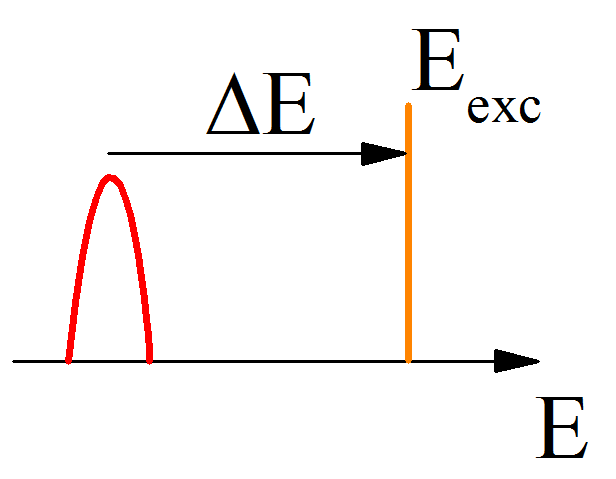}}
    \put(-230,198){\bf \textsf{b}} 
     \caption{
	Degree of circular polarization $\rho$ of the \textit{A} exciton of \mos{} (a) and \textit{A}$^-$ trion of \mose{} (b) plotted as a function of the 
photon excess energy $\Delta E$ (red and blue symbols). 
	Data and model from Ref.\,[\onlinecite{KioseoglouSciRep2016}], and data from Refs.\,[\onlinecite{Baranowski2017,Lagarde2014}] are displayed 
for comparison. Further comparison to the data referenced within Ref.\,[\onlinecite{KioseoglouSciRep2016}] is given in the Supplemental Material~\cite{SI}. 
Photoluminescence spectra are plotted as a guidance for the energy scale, superimposed Raman lines are marked with an asterisk (*). For \mose{}, a fit 
following $a/\sqrt{\Delta E}$ is included (dashed line). The inset depicts the definition of the excess energy $\Delta E$. 
	      }
     \label{fig:rho-excess-energy}
    \end{figure*}
            
      In Fig.~\ref{fig:rho-excess-energy} we plot the resonance behavior of the DOP $\rho$ (red and blue symbols). When increasing the excitation energy, thus providing excess energy $\Delta E$~\footnote{Note that we use the term "excess energy" for the energy difference of the excitation and observed emission. This should not be confused with, e.g., the exciton kinetic energy.} to the system, we observe a short plateau with full conservation ($\rho=1$) in the case of \mos{}. With further increasing excess energy, a decline of the DOP is observed. The energy range of the observed plateau is sample dependent [cf. Fig.~\ref{fig:rho-excess-energy}\,(a)], which we attribute to environmental effects, in particular surface adsorbates. The measurements of series 1 were taken ten months after exfoliation; other experiments with low laser power and in the cryogenic environment had been performed before. During this period the flake was stored in ambient conditions. In contrast, the flake for series 2 was put in vacuum and was measured shortly after exfoliation. 
 
      For \mose{}, we observe no plateau, but only the decline of polarization for rising excess energies $\Delta E$ [see Fig.~\ref{fig:rho-excess-energy}\,(b)]. 
      For comparison, data of Refs.~[\onlinecite{KioseoglouSciRep2016,Lagarde2014,Baranowski2017}] have been included.
      Note that the excess energy is not necessarily equivalent to the center-of-mass kinetic energy of the exciton, due to momentum conservation. 

      Considering the initial injection of carriers in the respective \textit{K} point $\rho_{\text{inj}}$ \i{}, the exciton lifetime $t_0$ \ii{}, and the 
      depolarization time $t_{\text{pol}}$ \iii{}, the DOP $\rho$ in steady state conditions can be written as \cite{Meier1984}
      \begin{equation}
	\rho=\frac{\rho_{\text{inj}}}{\left(1+t_0/t_{\text{pol}}\right)}.
      \end{equation}
We now discuss the potential effects of the change in excitation energy on the DOP. 
      \i{} The dependence on $\rho_{\text{inj}}$ reflects how many 
of the excited excitons (or more general charge carriers) are populating the \textit{K} valley corresponding to the circular polarization of the exciting 
photon. 
      By exciting at energies above the \textit{A} exciton states, processes potentially reducing the injected polarization become accessible, e.g., the 
absorption into the excited states of the \textit{A} exciton, the \textit{B} exciton state, and into the numerous contributions to the 
\textit{C} band\,\cite{Gillen2017}. 
Direct generation of the excited states of the exciton, e.g. in the $2s$ state at an excess energy of about 200\,meV\,\cite{Gillen2017,Hill2015} (i.e. for 
\mos{} at higher energies than the \textit{B} exciton), would contribute to a high initial polarization $\rho_{\text{inj}}$.
Generation of \textit{B} excitons in the same valley is allowed as the selection rules are the same; however, the electron and hole spin are opposite to the \textit{A} 
exciton [see Fig.~\ref{fig:intro}\,(b)]. Intravalley relaxation into the \textit{A} state would consequently 
need the spins to flip, but would result in a high $\rho_{\text{inj}}$. However, (potentially efficient single phonon) scattering 
from the \textit{B} state to an \textit{A} exciton state in the vicinity of the \textit{K'} valley would reduce $\rho_{\text{inj}}$. This process should be 
observable as a drop in DOP, when getting in resonance with the \textit{B} exciton. $\rho$, however, is already near zero at $\Delta E=150\,$meV. 
      
      
      In our experiment, we focus on the dynamics at excess energies below the spin-orbit splitting of the valence band, to investigate the \textit{A} 
exciton and trion depolarization mechanism not involving the above processes. 
Note that the strongest reduction of the DOP is clearly observable at energies below a relevant contribution of the \textit{B} exciton.
      
      \ii{} The intrinsic recombination time of the excitons depends linearly on the effective 
excitonic temperature\,\cite{Andreani1991} and thereby on the interplay of excess energy provided by the photons and heat transport within the sample and the 
cryostat. There are manifold extrinsic influences on the recombination time, of which the discussion is beyond the scope of this letter.
      
      \iii{} $t_{\text{pol}}$ reflects all processes reducing the DOP, such as the aforementioned depolarization mechanisms, which we now discuss in more 
detail. 
     
      In previous studies performed by Kioseoglou \etal{}\,\cite{KioseoglouAPL2012,KioseoglouSciRep2016,KioseoglouPSSRRL2016} (combination of variation of temperature and excitation energy to vary $\Delta E$), a similar decrease of the DOP in 
\mos{} and \mose{} was observed (cf. Fig.~\ref{fig:rho-excess-energy}, squares). Combining their own data and selected values from the 
literature (cf. Supplemental Material~\cite{SI}), they proposed a model of simultaneous scattering of electron and hole by two LA phonons between the \textit{K} and \textit{K'} points. In their model, scattering is only possible if sufficient energy is provided to emit the two LA 
phonons. These two phonons are phonons from the \textit{K} point, i.e., phonons with a wave vector $\overline{\Gamma K}$ [cf. Fig.~\ref{fig:intro}\,(a)]. However, 
the phonon dispersion of \mose{} is not experimentally known at the \textit{K} point. Therefore, the phonon energy of $E_{2\text{LA}}=39$\,meV can only be 
taken from 
calculations\,\cite{Horzum2013}. The same holds for \mos{}; however, very recently, experimental data have been obtained by inelastic x-ray scattering\,\cite{Tornatzky2018b}. These measurements find $E_{2\text{LA}}=59.6$\,meV, while density-functional theory calculations find 58.5\,meV\,\cite{Tornatzky2018b}.
In our measurements of \mos{}, on the other hand, the decrease of the DOP $\rho$ starts 
already well below 60\,meV excess energy [Fig.~\ref{fig:rho-excess-energy}\,(a)]. For \mose{}, no full conservation of polarization can be observed at all. 
Note that the measurement on \mose{} closest to resonance  was performed with an excess energy of 3\,meV, well below the energy of two LA phonons 
[Fig.~\ref{fig:rho-excess-energy}\,(b)].

      A further requirement to be met following the model of Ref.\,[\onlinecite{KioseoglouSciRep2016}] to access allowed states in the respective other 
\textit{K} 
valley is a spin flip of both the electron and the hole. In Ref.\,[\onlinecite{KioseoglouSciRep2016}] this is suggested to be either mediated by short range 
scattering on impurities or scattering through the nearly spin-degenerate $\Gamma$ point\,\cite{Mai2013}.

      The often observed localized exciton peak $L$ around $1.8$\,eV might hint at exciton scattering on defects.
      These could support the scattering between the \textit{K} valleys. Depending on the type of defect, it could, e.\,g., allow a spin flip and provide the 
necessary amount of momentum for an electron (hole) to scatter to the respective other \textit{K} valley, reducing the required excess energy to one 
LA(\textit{K}) phonon equivalent. 
      On the other hand, Mak \etal{}\,\cite{Mak2012a} have also observed a "trapped exciton" feature around 1.8\,eV and still achieved full conservation of 
circular polarization with approximately 39\,meV excess energy.

      The second spin flip mechanism, however, is not possible considering excitons, as no excitonic 
states in an accessible energy range are present. In the single-particle 
picture (assuming that the exciton binding energy somehow can be overcome), scattering through $\Gamma$ is not feasible for electrons as the states are about 
1\,eV higher than the conduction band minimum at the \textit{K} point. For the hole scattering through $\Gamma$ would require another LA(K) phonon to scatter 
from the $\Gamma$ to the \textit{K'} point, pushing the onset of the decline in polarization to $3E_{\text{LA}}$ [cf. Fig.~\ref{fig:intro}\,(a), dashed 
lines].
Further, also the single-layer valence bands at $\Gamma$ are at lower energies than accessible for the photogenerated hole by emission of a phonon\,\cite{Gillen2017}, hence, rendering them inaccessible without 
further 
energy transfer from the lattice.
     
      Another approach attributed the decline in the DOP $\rho$ with rising excess energy to the valley exchange mechanism\,\cite{Baranowski2017}. 
      The PLE measurements of Ref.\,[\onlinecite{Baranowski2017}] (at 4.5\,K, with varying excitation energy) for \mose{} are shown in 
Fig.~\ref{fig:rho-excess-energy}\,b) for comparison (stars). The 
authors present a rate equation model, where in steady-state conditions the maximum DOP $\rho$ depends on the times of excitonic recombination, scattering 
between the bright and dark states within the valley, and the valley exchange\,\cite{Baranowski2017}. From these they conclude a maximum possible polarization 
of about 35\,\% in \mose{}. 
      In our measurements, however, we find a higher degree of polarization.
      
      According to the  Maialle-Silva-Sham mechanism\,\cite{Maialle1993}, underlying the valley exchange, the exchange efficiency is proportional to the 
exciton center-of-mass momentum\,\cite{Yu2014}. 
      Assuming parabolic bands, the dependence of the conservation of polarization translates into a proportionality of $\rho\propto 1/\sqrt{\Delta E}$. 
This relation is fitted to our measurements of \mose{} [cf. Fig.~\ref{fig:rho-excess-energy}\,(b), dashed line] and is in good agreement with our data and the 
values acquired by Baranowski \etal{}\,\cite{Baranowski2017}, however deviates from those of Ref.\,[\onlinecite{KioseoglouSciRep2016}]. As the assumption of 
parabolic bands is only valid close to the band extrema, it is reasonable that the fit does not describe the trend at higher excess energies.
    States with $k_X>k_{\text{photon}}$, leading to the enhanced valley exchange, require an interaction with the lattice to obey momentum conservation. 
Momentum can either be gained by intravalley scattering of the photogenerated exciton on phonons or potentially by simultaneous generation (or 
annihilation) of a phonon. Successive (intravalley) cooling then enables the exciton in the $K'$ valley to again couple to light and relax radiatively.
    Consequently, this sequence of processes needs to be as fast as the emergence of the emission with opposite polarization to be relevant. In 
time-resolved Faraday rotation measurements on \mos{}, a fast decay with a time constant of 200\,fs has been measured\,\cite{DalConte2015a}.
Considering fast creation of hot excitons through thermalization with phonons on the order of 100\,fs\,\cite{Selig2018}, instantaneous valley 
exchange\,\cite{Schmidt2016} and fast thermalization (cooling into the light cone) in the \textit{K'} valley (again $\approx100$\,fs), one certainly is within 
the argued time regime.
    
      Intriguingly, the $1/\sqrt{\Delta E}$ dependence does not describe the acquired polarization values of \mos{}. 
      We therefore suggest that the difference might be attributed to different intrinsic properties of the materials. The interplay of different exciton- or 
electron-phonon coupling and exchange constants as well as lifetimes, and the availability of scattering or relaxation channels could lead to a different 
dominant depolarization mechanism.


      To conclude, we have presented PLE measurements of circularly polarized light emission from \mos{} and \mose{}. 
      We observe a high degree of conservation of circular polarization when exciting close to the resonance with the optical transition. However, the DOP 
$\rho$ is reduced as soon as the resonance condition is left. For \mose{}, we observe a degree of polarization $\rho$ of 84\,\% (79\,\%) for the 
\textit{A} exciton (trion) emission, which are the highest reported values for \mose{}, to the best of our knowledge. We find that the nature of the 
generation and valley depolarization of excitons in TMDCs is so far not fully understood, as none of the proposed mechanisms can describe the DOP in TMDCs in 
the vicinity of the resonances. However, the trends of the DOP of \mos{} and \mose{} are partly in agreement with the two described models stemming from 
the 2LA mechanism and the valley exchange, respectively. This suggests that more than one mechanism reduces the DOP in different TMDCs.
      Further investigations, especially of the electron and hole scattering involving phonons and defects, are needed for better understanding of the 
processes relevant for valleytronics.

  \section{Methods}
  \i\emph{Sample preparation.} 
  Single-layer samples were exfoliated by mechanical cleavage onto silicon with a top layer of 90\,nm silicon oxide, yielding flakes 
of $\approx10\,\mu$m size. 
  Single-layer flakes were identified by RGB contrast measurements with an optical microscope\,\cite{Li2013} and verified by the absence of the indirect gap 
transition (\mos) or of the $A_{1\text{g}}$/$A'_1$/$B^1_{2g}$ few-layer or bulk Raman mode\,\cite{Scheuschner2015} (\mose) ({cf.} section samples in the Supplemental Material). 

    \ii\emph{Optical measurements.}
 Microphotoluminescence spectra in backscattering geometry were acquired using a Dilor XY800 triple monochromator setup equipped 
with a liquid nitrogen cooled CCD. 
  For continuous wave excitation Rhodamine 6G (LC5900) and DCM special (LC6501) dye lasers as well as a Ti:sapphire laser were used. 
  The lasers were guided through a linear polarization filter before entering the microscope. Circular polarization was achieved by using an achromatic 
quarter-wave plate mounted just above the objective lens to avoid optics-induced distortions. The emitted circularly polarized luminescence was changed to 
linear polarization with the same quarter-wave plate, while selection of the respective polarization was performed by a linear analyzer. The power density on 
the sample was always kept below $10^8\,\text{W}/\text{m}^2$ (260\,$\mu$W on a $\approx2\,\mu$m laser spot) to avoid damage and excessive heating (cf. power 
series in the Supplemental Material).

  The samples were cooled to 20\,K by using a helium cooled cold-finger microcryostat.
  The sample temperature was kept at cryogenic temperatures during the time of the PLE measurements to avoid temperature-cycle-induced changes from, e.g., 
changing the dielectric environment due to absorbed water, damage induced by freezing adsorbates, etc.

%

\bigskip

\begin{acknowledgments}
{This work was supported in part by the Deutsche Forschungsgemeinschaft (DFG) within the Cluster of Excellence "Engineering of Advanced Materials" (Project No. EXC 315) (Bridge Funding) and the Collaborative Research Centre SFB787.}
We thank Axel Hoffmann (TU Berlin) for helpful discussions. 
\end{acknowledgments}



\end{document}